\documentclass[11pt]{article}

\usepackage{amsmath}
\usepackage{amssymb}

\newcommand{\fta}{{\sc fta}}
\topmargin = - 0.5 cm \textheight = 23 cm \textwidth = 15 cm
\begin{document} 
\pagenumbering{arabic} \setlength{\unitlength}{1cm}\cleardoublepage
\thispagestyle{empty}
\title{The Fine-Tuning Argument\thanks{To appear in \emph{The Challenge of Chance}, eds.\ Klaas Landsman \&\ Ellen van Wolde (Springer-Verlag, Heidelberg, 2016). Written for a general audience.}}
\author{Klaas Landsman\thanks{Department of Mathematical Physics,
Institute for Mathematics, Astrophysics, and Particle Physics, Faculty of Science, Radboud University Nijmegen, email \texttt{landsman@math.ru.nl}.}} \date{\today}
\maketitle
 \begin{abstract} 
\noindent 
Our laws of nature and our cosmos appear to be delicately fine-tuned for life to emerge, in a way that seems hard to attribute to chance.
In view of this, some have 
 taken the opportunity to revive the scholastic Argument from Design, whereas others have felt the need to explain 
 this apparent fine-tuning of the clockwork of the Universe by proposing the existence of a `Multiverse'.
We analyze this issue from a sober perspective.
Having reviewed the literature and having added several observations of our own, we conclude that  cosmic fine-tuning supports neither Design nor a Multiverse, since both of these fail at an explanatory level as well as in a more quantitative context of Bayesian confirmation theory (although there might be  
other reasons to believe in these ideas, to be found in religion and in inflation and/or  string theory, respectively).
In fact, fine-tuning and Design even seem to be at odds with each other, whereas   
the  inference from fine-tuning to a Multiverse only works if the latter is underwritten by an additional metaphysical hypothesis we consider unwarranted.
Instead, we suggest  that fine-tuning requires no special explanation at all, since it is not the Universe that is fine-tuned for life, but life that has been fine-tuned  to the Universe. 
\end{abstract}
\maketitle
\begin{center}\textbf{Motto}
\begin{quote}
A mild form of satire may be the appropriate antidote. Imagine, if you will, the wonderment of a species of mud worms who discover that if the constant of thermometric conductivity of mud were different by a small percentage they would not be able to survive. 
\hfill (Earman, 1987, p.\ 314)
\end{quote}
\end{center}
\newpage
\section{Introduction}
Twentieth Century physics and cosmology have revealed an astonishing path towards our existence, which appears to be 
predicated on a delicate interplay between the three fundamental forces that govern the behavior of matter at very small distances 
and the long-range force of gravity. The former control chemistry and hence life as we know it, whereas 
the latter is responsible for the overall evolution and structure of the Universe. 
\begin{itemize}
\item If the state of the hot dense matter immediately after the Big Bang had been ever so slightly different, then the Universe would either have rapidly recollapsed, or would have expanded far too quickly into a chilling,  eternal void. Either way, there would have been no `structure' in  the Universe in the form of stars and galaxies. 
\item Even given the above fine-tuning, if any one of the three short-range forces  had been just a tiny bit different  in strength, or if the masses of some elementary particles had been a little unlike they are, there would have been no  recognizable chemistry in either the inorganic or the organic  domain. Thus there would have been no Earth, no carbon, et cetera, let alone the  human brains to study those.
\end{itemize}
Broadly, five different responses to the impression of fine-tuning  have been given:
\begin{enumerate}
\item \emph{Design:} updating  the scholastic  \emph{Fifth Way} of Aquinas (1485), the Universe 
has been fine-tuned  with the emergence of (human) life among its designated purposes.\footnote{\label{fn1} ``The Fifth Way is based on the directedness of things. We observe that some things which lack awareness, namely natural bodies, act for the sake of an end. This is clear because they always or commonly act in the same manner to achieve what is best, which shows that they reach their goal not by chance but because they tend towards it.  
Now things which lack awareness do not tend towards a goal unless directed by something with awareness and intelligence, like an arrow by an archer. Therefore there is some intelligent being by whom everything in nature is directed to a goal, and this we call `God'."
Translation in  Kenny (1969, p.\ 96), to whom we also refer for a critical review of Aquinas's proofs of the existence of God. It is a moot point whether the Fifth Way is really an example of the medieval Argument of Design, which Aquinas expresses elsewhere as: ``The arrangement of diverse things cannot be dictated by their own private and divergent natures; of themselves they are diverse and exhibit no tendency to form a pattern. It follows that the order of many among themselves is either a matter of chance or must be attributed to one first planner who has a purpose in mind." (Kenny, 1969, p.\ 116).  Everitt (2004)  distinguishes between the Argument \emph{to} Design and the Argument \emph{from} Order, respectively,  both of which may still be found in modern Christian apologists such as Holder (2004), Swinburne (2004), 
K\"{u}ng (2005), and Collins (2009),  rebutted by e.g.\ Everitt (2004) and  Philipse (2012).
It is clear from  his writings (such as the General Scholium in \emph{Principia}) that Isaac Newton supported the Argument from Design, followed by Bentley (1692). Throughout early modern science, the gradual `reading' of the `Book of Nature', seen as a second `book' God had left mankind next to the Bible, was implicitly or explicitly seen as a confirmation of Design (Jorink, 2010). 
 Paley (1802) introduced the famous watchmaker analogy obliterated by Dawkins (1986). See also Barrow \& Tipler (1986) and Manson (2003) for overviews of the Argument from Design. 
 }
 \item \emph{Multiverse:} the idea that our Universe is just one among innumerably many, each of which is controlled by different parameters in the (otherwise fixed) laws of nature. This seemingly outrageous idea is actually endorsed by some of the most eminent scientists in the world, such as Martin Rees (1999) and Steven Weinberg (2007). The underlying  idea was nicely explained by Rees in a talk in 2003, raising the analogy with 
`an `off the shelf' clothes shop: ``if the shop has a large stock, we're not surprised to
find one suit that fits. Likewise, if our universe is selected from a multiverse, its
seemingly designed or fine-tuned features wouldn't be surprising.'' (Mellor 2002).
 \item \emph{Blind Chance:} constants of Nature  and initial conditions have  arbitrary values, and it is just a matter of coincidence that their actual values turn out to enable life.\footnote{ In the area of biology, a classical book expressing this  position is Monod (1971).}
 \item \emph{Blind Necessity:}  the Universe could not have been made in a different way or order, yet
 producing life is not among its goals since it fails to have any  (Spinoza, 1677).\footnote{The most prominent  modern Spinozist was Albert Einstein:
 ``there are no \emph{arbitrary} constants of this kind; that is to say, nature is so constituted that it is possible logically to lay down such strongly determined laws that within these laws only rationally completely determined constants occur (not constants, therefore, whose numerical value could be changed without destroying the theory."
 (Einstein in Schilpp, 1949, p.\ 63).}
  \item \emph{Misguided:} the fine-tuning problem should be resolved by some appropriate  therapy.
\end{enumerate}
We will argue that whatever reasons one may have for supporting the first or the second option, fine-tuning should not be among them.
Contemporary physics makes it hard to choose between the third and the fourth option (both of which  seem to have supporters among physicists and philosophers),\footnote{The famous ending of \emph{The First Three Minutes} by the physicist Weinberg---``The more the universe appears comprehensible, the more it also appears pointless.''---could   be bracketed under either. }
 but in any case our own sympathy  lies with the fifth. 

First, however, we have to delineate the issue. The \emph{Fine-Tuning Argument}, to be abbreviated by \fta\ in what follows, claims that the present Universe (including the laws that govern it and the initial conditions from which it has evolved) permits life only because these laws and conditions take a very special form, small changes in which would make life impossible. 
 This claim is actually quite ambiguous, in (at least)  two directions. 
 \begin{enumerate}
\item The  \fta\ being counterfactual (or, in Humanities jargon, being `what if' or `alternate'  history), it should be made clear what exactly is variable. Here the range lies between raw Existence itself at one end (Rundle, 2004; Holt, 2012; Leslie \& Kuhn, 2013)
 and  fixed laws of nature and a Big Bang with merely a few variable parameters at the other (cf.\ Rees, 1999;  Hogan, 2000;  
 Aguirre, 2001; Tegmark et al., 2006). 
 
Unless one is satisfied with pure philosophical speculation, specific technical results are only available at the latter end, to which we shall therefore restrict the argument. 
 \item It should be made clear what kind of `life' the Universe is (allegedly) fine-tuned for, and also, to what extent the emergence of whatever kind of life is deemed merely possible (if only in principle, perhaps with very low probability), or likely, or absolutely  certain. For example, should we fine-tune just for the possible existence of self-replicating structures like {\sc rna} and {\sc dna},\footnote{See e.g.\ Smith \& Szathm\'{a}ry (1995) and Ward \& Brownlee (2000) for theories of the origin of life.}  or for  ``a planet where enough wheat or rice could be cultivated to feed several billion people" (Ward \& Brownlee, 2000, p.\ 20), or  for one where morally (or indeed immorally) acting rational agents emerge (Swinburne, 2004), perhaps even minds the like of Newton and Beethoven? 
  
It seems uncontroversial that at the lowest end, the Universe should exhibit some kind of order and structure in order to at least enable life, whereas  towards  the upper end it has  (perhaps unsurprisingly) been claimed that essentially a copy of our Sun and our Earth (with even the nearby presence of a big planet like Jupiter to keep out asteroids) is required, including oceans, plate tectonics and other seismic activity, and a magnetic field helping to stabilize the atmosphere (Ward \& Brownlee, 2000).\footnote{The conservatism---perhaps even lack of imagination---of such scenarios is 
 striking. But science-fiction movies such as \emph{Star Trek, Star Wars}, \emph{E.T.}, \emph{My Stepmother is an Alien} (not to speak of \emph{Emmanuelle, Queen of the Galaxy})
 hardly do better. Conway's \emph{Game of Life} suggests that initial complexity is not at all needed to generate complex structures, which may well include intelligent life in as yet unknown guise.}
 
For most of the discussion we go for circumstances favoring simple carbon-based life;
 the transition to complex forms of life will only play a role in discussing the fine-tuning of our solar system (which is crucial to some and just a detail to others).\footnote{Reprimanding the late Carl Sagan, who expected intelligent life to exist in millions of places even within our own Galaxy, Ward \& Brownlee (2000) claim that whereas this might indeed apply to the most basic forms of life, it is the move to complex (let alone intelligent) life that  is extremely rare (because of the multitude of special conditions required), perhaps having been accomplished only on Earth.}
\end{enumerate}
   According to modern cosmology based on the (hot) Big Bang scenario,\footnote{See e.g.\ Rees (1999),  Ellis (2007), and Weinberg (2008), at increasing level of technicality.}
    this means  that the Universe 
 must be sufficiently  old and structured so that at least galaxies and several generations of stars  have formed; this already
takes billions of years.\footnote{
 The  reason (which may be baffling on first reading)  is that in addition to the light elements formed in Big Bang nucleosynthesis 
 (i.e., about 75\% hydrogen and 25\% helium, with traces of other elements up to Lithium, see Galli \& Palla, 2013), the heavier elements 
 in  the Periodic Table  (many of which are necessary  for  biochemistry and/or the composition of the Earth and its atmosphere) 
were formed in stars, to be subsequently blown into the cosmos by e.g.\ supernova explosions. In that way, some of these elements eventually ended up in our solar system, where they are indispensable in constituting both the Earth and 
ourselves. See Arnett (1996) for a technical account and Ward \& Brownlee (2000) for a popular one.} The subsequent move to viable planets and life then takes roughly a similar amount of time, so that within say half an order of magnitude the current age of the Universe seems necessary to support life. In view of the expansion of the Universe, a similar comment could be made about its size, exaggerated as it might seem for the purpose of explaining life on earth. 
\section{Evidence for fine-tuning}
Thanks  to impressive progress in both cosmology and (sub)nuclear physics, 
over the second half of the 20th Century it began to be realized that the above scenario is predicated on seemingly exquisite fine-tuning of some of the constants of Nature and initial conditions of the Universe. We just give some of the best known and best understood cases here.\footnote{See Barrow \& Tipler (1986), Leslie (1989), Davies (2006), Ellis (1997),  and Barnes (2012) for further examples and more detailed references. Stenger (2011) and Bradford (2011, 2013) attempt to play down the  accuracies claimed of fine-tuning, whilst  Aguirre (2001) casts doubt on its limited scope. }

 One of the first examples was the  `Beryllium bottleneck' studied by Hoyle in 1951, which is concerned with the mechanism through which stars produce carbon and oxygen.\footnote{In order to make carbon, two 
 $\mbox{}^4\mathrm{He}$ nuclei must collide to form  $\mbox{}^8\mathrm{Be}$, 
  upon which a third  $\mbox{}^4\mathrm{He}$ nucleus must join so as to give  $\mbox{}^{12}\mathrm{C}$ (from which, in turn,  $\mbox{}^{16}\mathrm{O}$ is made by adding another  $\mbox{}^4\mathrm{He}$ nucleus). This second step must happen extremely quickly, since the $\mbox{}^8\mathrm{Be}$ isotope formed in the first step is highly unstable. Without the exquisitely fine-tuned energy level in $\mbox{}^{12}\mathrm{C}$ (lying at the $\mbox{}^8\mathrm{Be}+\mbox{}^4\mathrm{He}$ reaction energy)
   predicted by Hoyle, this formation process would be far too infrequent to explain the known cosmic abundances. Opponents of anthropic reasoning would be right in pointing out that these abundances as such (rather than their implications  for the possibility of human life) formed the proper basis for Hoyle's prediction.}
  This was not only a major correct scientific prediction based on `anthropic reasoning' in the sense that some previously unknown physical effect (viz.\ the energy level in question) \emph{had} to exist in order to explain some crucial condition for life; it involves dramatic fine-tuning, too, 
in that the nucleon-nucleon force must lie near its actual strength within about one part in a thousand 
 in order to obtain the observed abundances of carbon and oxygen, which happen to be the
right amounts needed for life (Ekstr\"{o}m et al., 2010).

Another well-understood example from nuclear physics is the mass difference between protons and neutrons, or, more precisely, between the down quark and the up quark (Hogan, 2000).\footnote{Quarks are subnuclear particles that come in six varieties, of which only the so-called `up' and `down' quarks are relevant to ordinary matter. A neutron consists of one up quark and two down quarks, whereas a proton consists of two up quarks and one down quark. The electric charges (in  units where an electron has charge -1) are 2/3 for the up quark and -1/3 for the down quark, making a neutron electrically neutral (as its name suggests) whilst giving a proton charge +1.
Atoms consist of nuclei (which in turn consist of protons and neutrons) surrounded by electrons, whose total charge exactly cancels that of the nucleus.} This mass difference is positive (making the neutron heavier than the proton); if it weren't, the proton would fall apart and there would be no chemistry as we know it. On the other hand, the difference can't be too large, for otherwise stars (or hydrogen bombs, for that matter) could not be fueled by nuclear fusion and stars like our Sun would not exist.\footnote{Technically, 
the fundamental `$pp$-reaction'
(i.e., proton + proton $\rightarrow$ Deuteron + positron + neutrino), which lies at the beginning of nuclear fusion, would go in the wrong direction.} Both require a fine-tuning of the mass difference by about 10\%. 

Moving from fundamental forces to initial conditions, the solar system seems fine-tuned for life in various ways, most notably in the distance between the Sun and the Earth: if this had been greater (or smaller) by at most a few precent it would have been too cold (or too hot) for at least complex life to develop. Furthermore, to that effect the solar system must remain stable for billions of years, and after the first billion years or so the Earth should not be hit by comets or asteroids too often. Both conditions are  sensitive to the precise number and configuration of the planets 
 (Ward \& Brownlee, 2000).

Turning from the solar system to initial conditions of our Universe, but still  staying safely within the realm of well-understood physics and cosmology, Rees (1999) and others have drawn attention to the fine-tuning of another cosmological  number called $Q$, which gives the size of inhomogeneities, or `ripples', in the early Universe and  is of the order $Q\sim 0.00001$, or one 
part in a hundred thousand.\footnote{
The Universe is approximately 13.7 billion years old (which is about three times as old as the Earth).
Almost 400.000 years after the Big Bang, the Universe (which had been something like a hot soup of elementary particles until then) became transparent to 
electromagnetic radiation (which in everyday life includes light as well as radio waves, but whose spectrum is much larger) and subsequently became almost completely dark, as it is now. The so-called cosmic microwave background ({\sc cmb}, discovered in 1964 by Penzias and Wilson), which still pervades the Universe at a current temperature of about 3K ($= -270^{\circ}$ C),  is a relic from that era. It is almost completely homogeneous and isotropic, except for the ripples in question, whose (relative) size is given by the parameter $Q$. This provides direct information about the inhomogeneities of the Universe at the time the {\sc cmb} was formed, i.e., when it was 400.000 years old.} This parameter is fine-tuned by a factor of about ten on both sides (Rees, 1999; Tegmark et al., 2006): if it had been less than a tenth of its current value, then no galaxies would have been formed (and hence no stars and planets). If, on the other hand, it had been more than ten times its actual value, then matter would have been too lumpy, so that there wouldn't be  any stars (and planets) either, but only black holes. Either way, a key condition for life would be violated.\footnote{As analyzed by the Planck Collaboration (2014), variations in the constants of Nature would also affect the value of $Q$, which is ultimately determined by the physics of the early Universe.
Hence its known value of $10^{-5}$ constrains such variations; in particular, the fine-tuning of $Q$  necessary for life in turn fine-tunes the
fine-structure constant $\alpha$ (which controls electromagnetism and light) to within 1\% of its value (1/137).}

The expansion of the Universe is controlled by a number called $\Omega$, defined as the ratio between the actual matter density in the Universe and the so-called critical density.  If $\Omega\leq 1$, then the Universe would expand forever, whereas $\Omega>1$ would portend a recollapse. Thus  $\Omega=1$ is a critical point.\footnote{Roughly, the physics behind this is that at small matter density the (literally `energetic') expansion drive inherited from the Big Bang beats the gravitational force, which tries to pull matter together.}
It is remarkable enough that currently $\Omega\approx 1$ (within a few percent); what is astonishing  is that this is the case at such a  high age of the Universe. Namely, for $\Omega$ to retain  its (almost)  critical value for billions of years, it must have had this value right from the very beginning to a precision of at least 55 decimal places.\footnote{If not, the expansion would either have been too fast for structures like galaxies to emerge, or too slow to prevent rapid recollapse due to gravity, leading to a Big Crunch (Rees, 1999). This fine-tuning problem is often called the \emph{flatness problem}, since the Universe is exactly flat (in the sense of Einstein's Theory of General Relativity) when $\Omega=1$ (otherwise it either has a spherical or a hyperbolic geometry). The fine-tuning problem for $\Omega$ is generally considered to be solved by the (still speculative) theory of cosmic inflation (Liddle \& Lyth, 2000; Weinberg, 2008),
but even if this theory is correct, it merely shifts the fine-tuning from one place to another, since the parameters in any theory of inflation
have to be fine-tuned at least as much as $\Omega$; Carroll \& Tam (2010)  claim this would even be necessary to ten million decimals. In addition, the flatness problem may not be a problem at all,   like the horizon problem (McCoy, 2015).  } 

This leads us straight to Einstein's  cosmological constant $\Lambda$, which he introduced into his theory of gravity  in 1917 in order to (at least theoretically) stabilize the Universe against contracting or expanding, to
 subsequently delete it in 1929 after Hubble's landmark observation of the expansion of the Universe (famously calling its introduction  his ``biggest blunder''). 
 Ironically,  $\Lambda$ made a come-back in 1998 as the leading theoretical explanation of the (empirical) discovery that the expansion of the Universe is currently accelerating.\footnote{The Physics Nobel Prize  in 2011
was awarded to Perlmutter, Schmidt, and Riess  for this discovery. The cosmological constant  $\Lambda$ can theoretically
account for this acceleration as some sort of an invisible driving energy. Thus reinterpreted as  `dark energy',  $\Lambda$ contributes as much as  70\% to the energy  density of the Universe and hence it is currently also the leading contributor to $\Omega$ (Planck Collaboration, 2015).
} For us, the point is that even the currently accepted  value of  $\Lambda$ remains very close to zero, whereas according to (quantum field) theory it should be about 55 (some even say 120) orders of magnitude larger (Martin, 2012). This is often seen as a fine-tuning problem, because some compensating mechanism must be at work to cancel its very large natural value with a precision of (once again) 55 decimal places.\footnote{The broader context of this is what is called the \emph{naturalness problem} in quantum field theory, first raised by the Dutch Nobel Laureate Gerard 't Hooft in 1980. His claim was that a theory is unnatural if some parameter that is expected to be large is actually (almost) zero, unless there is a symmetry enforcing the latter. This generates its own fine-tuning problems, which we do not discuss here; see Grinbaum (2012).}

The fine-tuning of all  numbers considered so far seems to be dwarfed by 
a knock-down \fta\  given by Roger Penrose (1979, 2004), who claims that in order to produce a Universe that even very roughly looks like ours,
its initial conditions (among some generic set) must have been fine-tuned with a precision of one to $10^{10^{123}}$, arguably the largest number ever conceived: all atoms in the Universe would not suffice to write it out in full.\footnote{The number called ``googol'' that the internet company Google has (erroneously) been named after is `merely' $10^{100}$; Penrose's number is even much larger than a one with googol many zeroes.} Penrose's argument is an extreme version of an idea originally due to Boltzmann, who near the end of the 19th Century argued that the direction of time is a consequence of the increase of entropy in the future but not in the past,\footnote{This is a technical way of saying that heat flows from hot bodies to cold ones, that milk combines with tea to form a homogeneous mixture, that the cup containing it will fall apart if it falls on the ground, etc.; 
in all cases the opposite processes are physically possible, but are so unlikely that they never occur.} which requires an extremely unlikely initial state (Price, 1997; Uffink, 2007; Lebowitz, 2008). However, 
 this kind of reasoning is as brilliant as it is  controversial (Callendar, 2004, 2010; Earman, 2006; Eckhardt, 2006; 
 Wallace, 2010;  Schiffrin \& Wald, 2012). More generally,  the more extreme the asserted fine-tuning is, the more adventurous the underlying arguments are (or so we think).

 To be on the safe side, the  fine-tuning of $\Omega$, $\Lambda$, and Penrose's initial condition should perhaps be ignored, leaving us with  the other examples, and a few similar ones not discussed here. But these should certainly suffice to make a case for fine-tuning that is  serious enough to urge 
the reader to at least make a bet on one the five options listed  above.
\section{General arguments}
Before turning to a specific discussion of the Design and the Multiverse proposals, we make a few  critical (yet impartial) remarks that put the \fta\ in perspective (see also Sober, 2004; Manson, 2009).
Adherents of the \fta\ typically use analogies like the following:
 \begin{itemize}
\item  Someone lays out a deck of 52 cards after it has been shuffled. If   the cards emerge in some canonical order (e.g., the Ace of Spades down to 2, then the Ace of Hearts down to 2, etc.), then, on the tacit assumption that each outcome is equally (un)likely,   this very particular outcome supposedly cannot have been due to `luck' or chance. 
 \item  Alternatively,  if a  die is tossed a large number of times and the number 6 comes up every time, one would expect the die to be loaded, or the person who cast it to be a very skillful con man. Once again, each outcome was \emph{assumed}  equally likely. 
\end{itemize}

First, there is an underlying  assumption in the \fta\ to the effect that the `constants' of Nature as well as the initial conditions 
of the Universe (to both of  which the emergence of life is allegedly exquisitely sensitive) are similarly variable. This may or may not be the case;  the present state of science is not advanced enough to decide between chance and necessity concerning  the laws of nature and the beginning of the Universe.\footnote{Our own hunch tends towards necessity, for reasons lying in constructive quantum field theory (Glimm \& Jaffe, 1987): it turns out to be extremely difficult to give a mathematically rigorous construction of elementary particle physics, and the value of the constants may be fixed by the requirement of mathematical existence and consistency of the theory. For example, the so-called scalar $\varphi^4$ theory is believed to be trivial in four (space-time) dimensions, which implies that the relevant `constant of Nature' must be zero
(Fernandez, Fr\"{o}hlich, \&  Sokal, 1992).  This is as fine a fine-tuning as anything!
Similarly, in cosmology the Big Bang (and hence the initial conditions for the Universe it gives rise to) actually seems to be an illusion caused by the epistemic fact that we look at the quantum world through classical glasses. In this case, the requirement that
cosmology as we know it must actually emerge in the classical limit of some quantum theory (or of some future theory replacing quantum mechanics) may well fix the initial conditions.
}

Second,  granted that the `constants' etc.\ are  variable in principle (in the sense that values other than the current ones preserve the existence and consistency of the theories in which they occur), it is quite unclear to what extent they can vary and which variations may be regarded as `small'; yet the \fta\ relies on the assumption that even `small' variations would block the emergence of life (Manson, 2000). In the absence of such information, it would be natural to assume that any (real,  positive as appropriate) value may be assumed, but in that case mathematical probabilistic reasoning (which is necessary for the \fta\ in order to say that the current values are `unlikely') turns out to be impossible (McGrew et al., 2001; Colyvan et al., 2005; Koperski, 2005).\footnote{Suppose some constant takes values in the real axis.
In the absence of good reasons to the contrary, any alternative value to the current one should have the same probability.
But there is no flat probability measure on the real numbers (or on any non-compact subset thereof). Even if there were such a measure, any finite  interval, however large or small, would   have measure zero, so that one could not even (mathematically) express the difference between some constant permitting life if it just lies within some extremely small bandwidth (as the \fta\ has it), or in some enormously large one (which would refute the \fta). 
}
 But also if  a large but finite number of values (per constant or initial condition) needs be taken into account, it is hard to assign any kind of probability to any 
  of the alternative values;  even the assumption that each values is equally likely seems totally arbitrary
 (Everitt, 2004; Norton, 2010). 
 
 Nonetheless, these problems may perhaps be overcome and in any case, for the sake of argument we will continue to use the metaphors opening this section. 
\section{Critiquing  the inference of Design from fine-tuning}
The idea that cosmic fine-tuning originates in design by something like an intelligent Creator fits into a long-standing Judeo-Christian tradition,
where both the Cosmos and biology were explained in that way.\footnote{
See footnote \ref{fn1} and refs.\ therein. 
Note that a fine-tuning intelligent Creator
is still a long shot from the Christian God whom 
 Holder (2004), Swinburne (2004), K\"{u}ng (2005), and  Collins (2009) are really after!} Now that biology has yielded to the theory of Evolution proposed by
Darwin and Wallace in the mid 19th Century,\footnote{The Dutch primatologist De Waal (2013) recently noted how reasonable Creationism  originally was: animals known to the population of the  Middle East (where Judeo-Christian thought originated) included camels etc.\ but no  primates,  and hence all living creatures appeared very different from mankind.} the battleground has apparently moved back to the cosmos. Also there, Design remains a vulnerable idea.\footnote{In this context, it is worth mentioning that the familiar endorsement of the Big Bang by modern Christian apologists  (see footnote \ref{fn1}) as a scientific confirmation of the creation story in \emph{Genesis} 1 seems wishful thinking based on a common mistranslation of its opening line as ``In the beginning God created the heavens and the earth'' (and similarly in other languages), whereas the original Hebrew text does not intend the ``beginning'' as an absolute beginning of time but rather as the starting point of the action expressed by the following verb, whilst ``created'' should have read ``separated'' (Van Wolde, 2009). 
More generally, the  world picture at the time of  writing of \emph{Genesis} was that of a disk surrounded by water, 
the ensuing creation story not being one of \emph{creatio ex nihilo}, but one in which God grounds the Earth by setting it on pillars. This led to a tripartite picture of the Cosmos as consisting of water, earth, and heaven. 
See also Van Wolde's contribution to this volume, as well as Noordmans (1934). The remarkable \emph{creatio ex nihilo} story introduced by the Church Fathers  therefore lacks textual support from the Bible.} For the sake of argument we do not question the coherence of the idea of an intelligent Creator as such, although such a spirit seems chimerical  (Everitt, 2004; Philipse, 2012). 

First, in slightly different ways  
 Smith (1993) and Barnes (2012) both made the point that  the \fta\ does not claim, or support the conclusion, that the present Universe is \emph{optimal} for intelligent life. Indeed, it hardly seems not to be: even granted all the fine-tuning in the world as well as the existence of our earth with its relatively favorable conditions (Ward \& Brownlee, 2000), evolution has been walking a tightrope so as to produce as much as jellyfish, not to speak of primates (Dawkins, 1996). This fact alone  casts doubt on the \fta\ as an Argument of Design, for surely a benign Creator would prefer a Universe optimal for life, rather than one that narrowly permits it? From a theistic perspective it would seem far more efficient to have a cosmic architecture that is robust for its designated goal. 

Second, the inference to Design from the \fta\ seems to rest on a decisive tacit assumption whose exposure sustantially 
weakens this inference (Bradley, 2001).
The cards analogy \emph{presupposes} that 
there was such a thing as a canonical order; if there weren't, then any particular outcome would be thought of in the same way and would of course be attributed to chance.  Similarly, the dice metaphor  \emph{presupposes} that it is special for  6 to come up every single time;  probabilistically speaking,
every other outcome would have been just as (un)likely as the given sequence of sixes.\footnote{Entropy arguments do not improve the case for Design.
It is true that although the probability of a sequence of all sixes is the same as the probability of any other outcome, the former \emph{becomes} special if we coarse-grain the outcome space by counting the number of sixes in a given long sequence of throws and record that information only. The outcome with sixes only then becomes  extremely unlikely, since it could only have occurred in one possible way, whereas outcomes with fewer sixes have multiple realizations (the maximum probability  occurring when the number of sixes is about one-sixth of the total number of throws). The point is that the very act of coarse-graining again  \emph{presupposes} that six is a special value.} An then again, 
in the case of independently tunable constants of Nature and/or initial conditions, one (perhaps approximate) value of each of these must first be marked with a special label like `life-permitting' in order for the analogy with cards or dice (and hence the appeal of the \fta) to work.  
The \fta\ is predicated on such marking, which already  \emph{presupposes} that life is  special. 

It is irrelevant to this objection  whether or not life is indeed special; the point is that  the assumption  that life be special has to be made \emph{in addition} to the \fta\ in order to launch the latter on track to Design.  But  the inference from the (assumed) speciality of life to Design hardly needs the \fta: even if all values of the constants and initial conditions would lead to a life-permitting Universe, those who think that life is special would presumably point to a Creator. In fact, both by the arguments recalled at the beginning of this section and those below, their case would actually be considerably stronger
than the \fta.
 
 In sum, fine-tuning is not by itself sufficient as a source for an Argument of Design; it is the \emph{combination} with an assumption to the effect that life is somehow singled out, preferred, or special. But that assumption is the one that carries the inference to Design; the moment one makes it, fine-tuning seems counter-productive rather than helpful.  

Attempts to give the Design Argument a quantitative turn (Swinburne, 2004; Collins, 2009)  make things even
 worse  (Bradley, 2002; Halvorson, 2014). Such attempt are typically based on \emph{Bayesian Confirmation Theory}.
This is a mathematical technique for the analysis and computation of 
the probability $P(H|E)$ that a given hypothesis $H$ is true 
in the light of certain evidence $E$ (which may speak for or against $H$, or may be neutral).
Almost every argument in Bayesian Confirmation Theory is ultimately based on \emph{Bayes' Theorem}
$$P(H|E)=P(E|H)\cdot P(H)/P(E),$$
 where $P(E|H)$ is the probability that $E$ is true given the truth of the hypothesis $H$, whilst $P(H)$ and $P(E)$ are the probabilities that $H$ and $E$ are true without knowing $E$ and $H$, respectively (but typically assuming certain background knowledge common to both $H$ and $E$, which is very important but has been suppressed from the notation).\footnote{
 This implies, in particular, that $P(H|E)>P(H)$, i.e., $E$ confirms $H$, if and only if $P(E|H)>P(E)$, which is often computable.
 The probabilities in question are usually (though not necessarily) taken to be \emph{epistemic} or (inter)\emph{subjective}, so that the whole discussion is concerned with probabilities construed as numerical measures of \emph{degrees of belief}. For technical as well as philosophical background on Bayesianism see e.g.\
 Howson \& Urbach (2006), Sober (2008), and Handfield (2012).
 }
 
  In the case at hand, theists want to argue that the Universe being fine-tuned for  \emph{L}ife makes  \emph{D}esign more likely, i.e., that $P(D|L)>P(D)$, or, equivalently, that $P(L|D)>P(L)$ (that is, Design favors life).  
 The problem is that theists do not merely 
 ask for the latter inequality; what they really believe is that $P(L|D)\approx 1$, for the existence of God should
 make the emergence of life almost certain.\footnote{Swinburne (2004), though, occasionally assumes that $P(L|D)=1/2$ as a subjective probability (based on our ignorance of God's intentions), which still makes his reasoning vulnerable to the argument below.  In Swinburne (2004), arguments implying $P(L|D)>P(L)$ are called $C$-\emph{inductive},
 whereas the stronger ones implying $P(L|D)> 1/2$ are said to be $P$-\emph{inductive}. Swinburne's  strategy is to combine a large number of  $C$-inductive arguments into a single overarching $P$-inductive one, but according to Philipse (2012) every single one of  Swinburne's $C$-inductive arguments is actually invalid (and we agree with Philipse).}
 For simplicity,  first assume that $P(L|D)=1$. Bayes' Theorem then gives $P(D|L)=P(D)/P(L)$, whence
 $P(D)\leq P(L)$. More generally, assume
$P(L|D)\geq 1/2$, or, equivalently, $P(L|D)\geq P(\neg L|D)$, where $\neg L$ is the proposition that life does not exist.
  If $(D,L)$ is the conjunction of $D$ and $L$, we then have
$$
P(D)=P(D,L)+P(D,\neg L)\leq 2P(D,L)\leq 2P(L),
$$
since  $P(D,\neg L)\leq P(D,L)$ by assumption. 
Thus a negligible prior probability of life  (on which assumption the \fta\ is based!) implies a hardly less negligible prior probability of Design. This inequality make the Argument from Design self-defeating as an explanation of fine-tuning, but in any case, both the interpretation and the numerical value of $P(D)$ are so obscure and ill-defined that the whole discussion seems, well, scholastic. 
\section{Critiquing the inference of a Multiverse from fine-tuning}
The idea of Design may be said to be human-oriented in a spiritual way, whereas the idea of a Multiverse more technically hinges on the existence of observers,  as expressed by the so-called (weak) \emph{Anthropic Principle} (Barrow \& Tipler, 1986; Bostrom, 2002). The claim is that there are innumerable Universes (jointly forming a `Multiverse'), each having its own `constants' of Nature and initial conditions, so that, unlikely as the life-inducing values of these constants and conditions 
in our Universe may be, they simply \emph{must}  occur within this unfathomable plurality. The point, then, is that we have to observe precisely those values because in other Universes there simply are no observers. This principle has been labeled both `tautological' and `unscientific'. Some love it and some hate it, but  we do not need to take sides in this debate: all we wish to do is find out whether or not the \fta\ speaks in favour of a Multiverse, looking  at both an explanatory and  a probabilistic level. Thus the question is whether the (alleged) fact of fine-tuning is (at least to some extent) explained by a Multiverse, or if, in the  context of Bayesian confirmation theory, the evidence of fine-tuning increases the probability of the hypothesis that a Multiverse exists.\footnote{ 
Although fine-tuning has been claimed  (notably by Rees, 1999) to provide independent motivation for believing in a Multiverse, the existence of a Multiverse may be a technical consequence of some combination of string theory (Susskind, 2005; Schellekens, 2013) and cosmological inflation (Liddle \& Lyth, 2000; Weinberg, 2008). Both theories are highly speculative, though (the latter less so than the former), and concerning the `landscape' idea it is hard to avoid the impression that string theorists turn vice into virtue by selling the inability of string theory to predict anything as an ability to predict everything (a similar worry may also apply to inflation, cf.\ Smeenk, 2014). Let us also note that even if it were to make any sense, the `emergent multiverse' claimed to exist in the so-called Many-Worlds (or Everett) Interpretation of quantum mechanics (Wallace, 2012) is a red herring in the present context, for, as far as we understand, all its branches have exactly the same laws of nature, including the values of the constants.}
To get the technical discussion going, the following metaphors have been used:
\begin{itemize}
\item  Rees's `off the shelf' clothes shop has already been mentioned in the Introduction: if someone enters a shop that sells suits in one size only (i.e., a single Universe), it would be amazing if it fitted (i.e., enabled life). However, if all sizes are sold (in a Multiverse, that is), the client would not at all be surprised to find a suit that fits.
\item  Leslie's (1989) 
 firing squad analogy states that someone should be executed by a firing squad, consisting of many marksmen, but they all miss. This amounts to fine-tuning for life in a single Universe. The thrust of the metaphor arises when the lucky executee is the sole survivor among a large number of other convicts, most or all of whom are killed (analogously to the other branches of the Multiverse, most or all of which are inhospitable to life). The idea is that although each convict had a small \emph{a priori} probability of not being hit,  if there are many of them these small individual probabilities of survival  add up to a large probability that \emph{someone} survives. 
\item Bradley (2009, 2012) considers an urn that is filled according to a random procedure:
\begin{itemize}
\item If a coin flip gives Heads (corresponds to a single Universe), 
either a small ball (life) or a large one (no life) is entered (depending on a further coin flip). 
\item In case of Tails (modeling a `Binaverse' for simplicity), two balls enter the urn, whose sizes depend on two further coin flips (leaving four possibilities). 
\end{itemize}
Using a biased drawing procedure that could only yield either a small ball or nothing, a small ball is obtained (playing the role of a life-enabling Universe). A simple Bayesian computation shows that this outcome confirms Tails for the initial flip. 
\end{itemize}
Each of these stories is insightful and worth contemplating. For example, the first one nicely contrasts the Multiverse with Design, which would correspond to bespoke tailoring and hence, at least from a secular point of view, commits the fallacy of putting the customer (i.e., life) first, instead of the tailor (i.e., the Universe as it is). The Dostoyevskian character of the second highlights the Anthropic Principle, whose associated selection effects  (Bostrom, 2002) are also quantitatively taken into account by the third. 

Nonetheless, on closer inspection each is sufficiently vulnerable to fail to clinch the issue in favour of the Multiverse. 
One point is that although each author is well aware of (and the second and the third even respond to) the \emph{Inverse Gambler's Fallacy}  (Hacking, 1987),\footnote{The
 \emph{Gambler's Fallacy} is the mistake that after observing say 35 throws of two fair dice without a double six, this preferred outcome has become more likely (than 1/36) in the next throw.} this fallacy is not really  avoided (White, 2000).
  In its simplest version, this is the mistake made by a gambler who enters a casino or a pub, notices that a double six is thrown at some table, and asks if this is the first roll of the evening (his underlying false assumption being that this particular outcome is more likely if many rolls preceded it). Despite claims to the contrary (Leslie, 1988;  Manson \& Thrush, 2003;  Bradley, 2009 \& 2012), Hacking's analysis that this is precisely the error made by those who favor a Multiverse based on the \fta\ in our opinion still stands. 
  For example, in Rees' analogy of the clothes shop, what needs to be explained is not that \emph{some} suit in the shop turns out to fit the customer, but that the one \emph{he happens to be standing in front of} does. Similarly, the probability that a \emph{given} executee survives is independent of whoever else is going to be shot in the same round. And finally, the relevant urn metaphor is not the one described above, but the one in which Tails leads to the filling of \emph{two} different urns with one ball each. 
Proponents of a Multiverse correctly state that its existence would increase the probability of life existing in \emph{some} Universe,\footnote{Bradley (2012, p.\ 164)  states \emph{verbatim} that he is computing the probability that ``At least one universe has the right constants for life'', other authors doing likewise either explicitly or tacitly.} but
this is only relevant to the probability of life in \emph{this} Universe if one  identifies  \emph{any} Universe with the same properties as ours with \emph{our} Universe.\footnote{Bradley (2009)  counters objections like Hacking's by the claim that ``if there are many Universes, there is a greater chance that Alpha [i.e., our Universe] will exist''. This implies the same identification.} Such an identification may be suggested by the (weak) Anthropic Principle,
 but its is by no means implied by it, and one should realize that the inference of a 
  Multiverse from the \fta\ implicitly hinges on this additional assumption.\footnote{We side with  Hartle \& Srednicki (2007) in believing that the identification in question is solidly wrong: 
 ``This notion presupposes that we exist separately from our physical description.  But  we are not separate from our physical description in our data;  we \emph{are} the physical system described (\ldots) 
  It is \emph{our data} that is used in a Bayesian analysis to discriminate between theories. What other hypothetical observers with data different from ours might see, how many of them there are, and what properties they might or might not share with us (\ldots) are irrelevant for this process."}

Moving from a probabilistic to an  explanatory context, we follow Mellor (2002) in claiming that if anything, a Multiverse would make fine-tuning even more puzzling. Taking the firing squad analogy,
there is no doubt that the survival of a single executee is unexpected, but the question is whether it may be explained (or, at least, whether it becomes less unexpected) by the assumption that simultaneously, many other `successful' executions were taking place. From the probabilistic point of view discussed above, their presence should have no bearing on the case of the lone survivor, whose luck remains as amazing as it was. From another, explanatory point of view, 
\emph{it makes his survival even more puzzling}, since we now know from this additional information about the other executions that apparently the marksmen usually do kill their victims. 
 \section{Conclusion}
Already the uncontroversial examples that feed the \fta\ suffice to produce the fascinating insight that the formal structure of our current theories of (sub)nuclear physics and cosmology (i.e., the Standard Model of particle physics and Einstein's theory of General Relativity) is insufficient to predict the rich phenomenology these theories give rise to: the precise values of most (if not all) constants and initial conditions play an equally decisive role. This is a recent insight:  even a physicist having the stature of Nobel Laureate Glashow (1999, p.\ 80) got this wrong,  having initially paraphrased the situation well:
\begin{quote}
``Imagine a television set with lots of knobs: for focus, brightness, tint, contrast, bass, treble, and so on. The show seems much the same whatever the adjustments, within a large range. \emph{The standard model  is a lot like that}.  

Who would care if the tau lepton mass were doubled or the Cabibbo
angle halved? The standard model has about 19 knobs. They are not really adjustable:
they have been adjusted at the factory. Why they have their values are 19 of the most
baffling metaquestions associated with particle physics.'' 
\end{quote}
In our view, the insight that  \emph{the standard model is not like that at all} is the real upshot of the \fta.\footnote{Callender (2004) understandably misquotes Glashow, writing:  
``The standard model is \emph{not} like that.''}
 Attempts to draw further conclusions from it in the direction of either \emph{Design} or a \emph{Multiverse} are, in our opinion, unwarranted. For one thing, as we argued,
 at best they  fail to have any explanatory or probabilistic thrust (unless they rely on precarious additional assumptions), and at worst fine-tuning actually seems to turn against them. 
 
Most who agree with this verdict would probably feel  left with a choice between the options of \emph{Blind Chance} and \emph{Blind Necessity};  the present state of science does not allow us to make such a choice now (at least not rationally), and the question  even arises if science will ever be able to make it  (in a broader context), expect perhaps  philosophically (e.g., \`{a} la Kant).  However, we would like to make a brief case for the fifth position, stating that the fine-tuning problem is \emph{misguided} and that all we need to do is to clear away confusion.  

There are analogies and differences between cosmic fine-tuning for life through the laws of Nature and the initial conditions of the Universe, as discussed so far, and Evolution in the sense of Darwin and Wallace. The latter is based on random (genetic) variation, survival of the fittest, and heritability of fitness. All these are meant to apply locally, i.e., to life on Earth. 
We personally feel that arguments to extend these principles to the Universe in the sense that the Cosmos may undergo some kind of `biological' evolution, having descendants born in singularities, perhaps governed by different laws and initial conditions (some of which, then,  might be `fine-tuned for life', as in the Multiverse argument), as argued by e.g.\ Wheeler (in Ch.\ 44 of Misner, Thorne, \&\ Wheeler, 1973) and Smolin (1997), imaginative as they may be,  are too speculative to merit serious discussion. Instead, the true analogy seems to be as follows: as far as the emergence and subsequent evolution of life are concerned, the Universe and our planet Earth should simply be taken as given. Thus the fundamental reason we feel `fine-tuning for life' requires no explanation is this:\footnote{
From this point of view, what we see as  the essential mistake made by those who feel fine-tuning does require an explanation is similar to what Butterfield (1931) famously christened `Whig History', i.e., ``the tendency in many historians to write on the side of Protestants and Whigs, to praise revolutions provided they have been successful, to emphasize certain principles of progress in the past and to produce a story which is the ratification if not the glorification of the present.'' (quoted from the Preface).}

\emph{Our Universe has not been fine-tuned for life: life has been fine-tuned  to our Universe. }
\subsection*{Acknowledgement}
The author is indebted to Jeremy Butterfield, Craig Callender,  Jelle Goeman, Olivier Hekster, Casey McCoy, Herman Philipse, Jos Uffink, and Ellen van Wolde for comments, discussions, and encouragement, all of which considerably improved this paper. 
\begin{small}
\section*{References}
\begin{trivlist}
\item Aguirre, A. (2001). The Cold Big-Bang cosmology as a counter-example to several anthropic arguments.
\emph{ Physical Review} D64, 083508.
\item Aquinas, Th. (1485). \emph{Summa Theologica}. Basel: M. Wenssler. Written 1268--1274.
\item Arnett, D. (1996). \emph{Supernovae and Nucleosynthesis: An Investigation of the History of Matter, From the Big Bang to the Present.} Princeton: Princeton University Press.
\item Barnes, L.A. (2012). The fine-tuning of the Universe for intelligent life. \emph{Publications of the Astronomical Society of Australia} 29, 529--56.
\item Barrow, J.D., Tipler, F. (1986). \emph{The Anthropic Cosmological Principle}. Oxford: Clarendon Press. 
\item Bentley, R. (1692). \emph{A Confutation of Atheism from the Origin and Frame of the World}. London:  Mortlock.
\item Bostrom, N. (2002). \emph{Anthropic Bias: Observation Selection Effects in Science and Philosophy}.
New York: Routledge. 
\item Bradford, R. (2011). The inevitability of fine tuning in a complex Universe.
\emph{International Journal of Theoretical Physics} 50, 1577--1601. 
\item Bradford, R. (2013). Rick's critique of cosmic coincidences. Website (consulted April 6th, 2015),\\ \texttt{www.rickbradford.co.uk/Coincidences.html}.
\item Bradley, D. (2009). Multiple Universes and the observation selection effects. \emph{American Philosophical Quarterly} 46, 61--72. 
\item Bradley, D. (2012). Four problems about self-locating belief. \emph{Philosophical Review} 121, 149--177.
\item Bradley, M.C. (2001). The fine-tuning argument. \emph{Religious Studies} 37, 451--466.
\item Bradley, M.C. (2002). The  fine-tuning argument: the Bayesian version. \emph{Religious Studies} 38, 375--404.
\item Butterfield, H. (1931). \emph{The Whig Interpretation of History}. London: Bell. 
\item Callender, C. (2004). Measures, explanations and the past: Should `special' initial conditions be explained?
\emph{The British Journal for the Philosophy of Science} 55, 195--217. 
\item Callender, C. (2010). The Past Hypothesis meets gravity. \emph{Time, Chance, and Reduction: Philosophical Aspects of Statistical Mechanics}, eds.\ Ernst, G., H\"{u}tteman, A., pp.\ 34--58. Cambridge: Cambridge University Press.
\item Carroll, S.M., Tam, H. (2010). Unitary evolution and cosmological fine-tuning.  \texttt{arXiv:1007.1417}.
\item Collins, R. (2009). The teleological argument: An exploration of the fine-tuning of the Universe. \emph{The Blackwell Companion to Natural Theology}, eds.\ Craig, W.L. \& Moreland, J.P.,
pp.\ 202--281. Chichester: Wiley--Blackwell.
\item Colyvan, M., Garfield, J.L., Priest, G. (2005). Problems with the argument from fine tuning.
\emph{Synthese} 145, 325--338. 
\item Davies, P. (2006). \emph{The Goldilocks Dilemma: Why is the Universe Just Right for Life?}
 New York: Mariner.
\item Dawkins, R. (1986). \emph{The Blind Watchmaker}. New York: Norton.
\item Dawkins, R. (1996). \emph{Climbing Mount Improbable}. New York: Norton.
\item De Waal, F. (2013). \emph{The Bonobo and the Atheist}. New York: Norton.
\item 
Earman, J. (1987). The SAP also rises: A critical examination of the anthropic principle. \emph{American Philosophical Quarterly} 24, 307--317.
\item 
Earman, J. (2006). The ``Past Hypothesis'': Not even false.  \emph{Studies in History and Philosophy of Modern Physics} 37, 399--430.
\item Eckhardt, W. (2006). Causal time asymmetry. \emph{Studies in History and Philosophy of Modern Physics} 37, 439--466.
\item Ekstr\"{o}m, S. et al. (2010). 
Effects of the variation of fundamental constants on Population {\sc iii} stellar evolution.
\emph{Astronomy \& Astrophysics} 514, A62.
\item Ellis, G.G.R. (2007). Issues in the philosophy of cosmology. \emph{Philosophy of Physics Part B}, eds.\ Butterfield, J. \& Earman, J.,
pp.\ 1183--1286. Amsterdam: Elsevier. 
\item Everitt, N. (2004).  \emph{The Non-Existence of God}. London: Routlegde.
\item   Fernandez, R.,  Fr\"{o}hlich, J., Sokal, A.D. (1992). \emph{Random Walks, Critical Phenomena, and Triviality in Quantum Field Theory}. Heidelberg:  Springer.
\item Glimm, J., Jaffe, A. (1987). \emph{Quantum Physics: A Functional Integral Point of View}.
New York: Springer. 
\item Galli, D., Palla, F. (2013). The dawn of chemistry. \emph{Annual Review of Astronomy and Astrophysics} 51, 163--206.
\item Glashow, S.L (1999). Does quantum field theory need a foundation? \emph{Conceptual Foundations of Quantum Field Theory}, ed. Cao, T.Y., pp.\ 74--88. Cambridge: Cambridge  University Press. 
\item Grinbaum, A. (2012). Which fine-tuning problems are fine? \emph{Foundations of Physics} 42, 615--631. 
\item Hacking, I. (1987). The inverse gambler's fallacy: The argument from design. The Anthropic Principle applied to Wheeler Universes.
\emph{Mind} 96, 331--340.
\item Halvorson, H. (2014). A probability problem in the fine-tuning argument. \\
\texttt{philsci-archive.pitt.edu/11004/}.
\item Hartle, J.B., Srednicki, M. (2007). Are We Typical? \emph{Physical Review D} 75, 123523. 
\item Handfield, T. (2012). \emph{A Philosophical Guide to Chance}. Cambridge: Cambridge  University Press. 
\item Hogan, C.J. (2000). Why the universe is just so.  \emph{Reviews of Modern Physics} 72, 1149--1161.
\item Holt, J. (2012). \emph{Why Does the World Exist? An Existential Detective Story}. New York: W.W. Norton.
\item Howson, C., Urbach, P. (2006). \emph{Scientific Reasoning: The Bayesian Approach}. Third Edition.
Chicago: Open Court. 
\item Jorink, E. (2010). \emph{Reading the Book of Nature in the Dutch Golden Age, 1575--1715}. Leiden: Brill.
\item Kenny, A. (1969). \emph{The five ways: St Thomas Aquinas' proofs of God's existence}. London: Routledge \& Kegan Paul.
\item Koperski, J. (2005). Should we care about fine-tuning?
\emph{The British Journal for the Philosophy of Science} 56, 303--319. 
\item K\"{u}ng, H. (2005). \emph{Der Anfang aller Dinge: Naturwissenschaft und Religion}. M\"{u}nchen: Piper. 
\item Lebowitz, J.L. (2008). From time-symmetric microscopic dynamics to time-asymmetric macroscopic behavior: An overview. \emph{Boltzmann's Legacy}, eds.\ Gallavotti, G., Reiter, W.L., Yngavson, J., pp.\ 38--62. 
Z\"{u}rich: European Mathematical Society. 
\item Leslie, J. (1988).  No Inverse Gambler's Fallacy in Cosmology. \emph{Mind} 97,  269--272.
\item Leslie, J. (1989). \emph{Universes}. London: Routledge. 
\item Leslie, J., Kunh, R.L. (2013). \emph{The Mystery of Existence: Why is there Anything at All?}. Chichester: Wiley--Blackwell.
\item Liddle, A.R., Lyth, D.A. (2000). \emph{Cosmological Inflation and Large-scale Structure}.
Cambridge: Cambridge  University Press. 
\item Manson, N.A. (2000). There is no adequate definition of `fine-tuned for life'. \emph{Inquiry} 43, 341--352.
\item Manson, N.A., ed. (2003). \emph{God and Design}. London: Routledge. 
\item Manson, N.A. (2009). The fine-tuning argument. \emph{Philosophy Compass} 4, 271--286.
\item  Manson, N.A., Thrush, M.J. (2003). Fine-tuning, multiple Universes, and the ``this Universe'' objection.
\emph{Pacific Philosophical Quarterly} 84, 67--83.
\item Martin, J. (2012). Everything you always wanted to know about the cosmological constant problem (but were afraid to ask).
\emph{Comptes Rendus Physique}
 13, 566--665.
 \item McCoy, C. (2015). What is the horizon problem? Preprint. 
 \item  McGrew, T., McGrew, L., Vestrup, E. (2001). Probabilities and the Fine-Tuning
Argument: a Sceptical View. \emph{Mind} 110, 1027--1038.
\item Mellor, D.H. (2002). Too many Universes. \emph{God and Design: The Teleological Argument and  Modern Science}, ed.\ Manson, N.A., pp.\ 221--228. London: Routledge. 
\item Misner, C., Thorne, K., Wheeler, J.A. (1973). \emph{Gravitation}. San Francisco: Freeman. 
 \item Monod, J. (1971). \emph{Chance and Necessity: An Essay on the Natural Philosophy of Modern Biology}.
 New York: A. Knopf.
  \item Noordmans, O. (1934). \emph{Herschepping}. Zeist: De Nederlandsche Christen-Studenten-Vereeniging.
  \item Norton, J.D. (2010). Cosmic confusions: Not supporting versus supporting not.
 \emph{Philosophy of Science} 77, 501--523.
\item Paley, W. (1802). \emph{Natural Theology: Or Evidences of the Existence and Attributes of the Deity Collected from the Appearances of Nature}. London: J. Foulder.
\item Penrose, R. (1979). Singularities and time-asymmetry. \emph{General Relativity: An Einstein Centenary Survey}, eds.\ Hawking, S.W., Israel, W.,
pp.\ 581--638.
\item Penrose, R. (2004). \emph{The Road to Reality: A Complete Guide to the Laws of the Universe}. London: Jonathan Cape. 
\item Philipse, H. (2012). \emph{God in the Age of Science? A Critique of Religious Reason}. Oxford: Oxford  University Press. 
\item Planck Collaboration (2014). \emph{Planck} intermediate results. {\sc xxiv}. Constraints on variation of fundamental constants.
\emph{Astronomy \& Astrophysics}, in press. \texttt{arXiv:1406.7482}. 
\item Planck Collaboration (2015). Planck 2015 results. XIII. Cosmological parameters.  \texttt{arXiv:1502.01589}.
\item Price, H. (1997). \emph{Time's Arrow \& Archimedes' Point}. Cambridge: Cambridge  University Press. 
\item Rees, M. (1999). \emph{Just Six Numbers}. London: Weidenfeld \& Nicolson. 
\item Rundle, B. (2004). \emph{Why There Is Something Rather Than Nothing}. Oxford: Oxford  University Press. 
\item Schellekens, A.N. (2013). Life at the interface of particle physics and string theory. \emph{Reviews of Modern Physics}
85, 1491--1540.
\item Schiffrin, J.S., Wald, R.M. (2012).  Measure and probability in cosmology. \emph{Physical Review D}  86, 023521.
\item Schilpp, P.A. (1949). \emph{Albert Einstein: Philosopher-Scientist}, Vol.\ 1. La Salle: Open Court. 
\item Smeenk, C. (2014). Predictability crisis in early universe cosmology. \emph{Studies in History and Philosophy of Modern Physics}
46, 122--133. 
\item Smith,, J.M., Szathm\'{a}ry, E. (1995).\emph{The Major Transitions in Evolution}. Oxford: Freeman. 
\item Smith, Q. (1993).  \emph{Theism, Atheism and Big Bang Cosmology} (with Craig, W.L.), pp.\ 203--204.
Oxford: Clarendon Press. 
\item Smolin, L. (1997). \emph{The Life of the Cosmos}.  Oxford: Oxford  University Press. 
\item Sober, E. (2004). The Design Argument. \emph{Blackwell Guide to the Philosophy of Religion}, ed.\ 
Mann, W.E., pp. 117--147. Oxford: Blackwell.
\item Sober, G. (2008). \emph{Evidence and Evolution}. Cambridge: Cambridge  University Press. 
\item Spinoza, B. de (1677). \emph{Ethica Ordine Geometrico Demonstrata}. Amsterdam: J. Rieuwertsz.
\item Stenger, V. (2011). \emph{The Fallacy of Fine-Tuning: Why the Universe Is Not Designed For Us}.
Amherst (NY): Prometheus. 
\item Susskind, L. (2005). \emph{The Cosmic Landscape: String Theory and the Illusion of Intelligent Design}.  New York: Little, Brown.
\item Swinburne, R. (2004). \emph{The Existence of God}. Second Edition. Oxford: Oxford  University Press. 
\item  Tegmark, M.,  Aguirre, A., Rees, M.J.,  Wilczek, F. (2006). Dimensionless constants, cosmology, and other dark matters. \emph{Physical Review} D 73, 023505. 
\item Uffink, J. (2007). Compendium of the foundations of classical statistical physics. 
 \emph{Philosophy of Physics Part B}, eds.\ Butterfield, J. \& Earman, J.,
pp.\ 923---1074. Amsterdam: Elsevier. 
 \item Van Wolde, E. (2009). Why the verb $b\overline{a}r\overline{a}'$ does not mean `to create' in Genesis 1.1-2.4a.
\emph{Journal for the Study of the Old Testament} 34, 3--23
\item Wallace, D. (2010). Gravity, entropy, and cosmology: In search of clarity.
\emph{The British Journal for the Philosophy of Science} 61, 513--540. 
\item Wallace, D. (2012). \emph{The Emergent Multiverse: Quantum Theory According to the Everett Interpretation}.
 Oxford: Oxford  University Press. 
 \item Ward, P., Brownlee, D. (2000). \emph{Rare Earth}. New York: Copernicus Books. 
 \item Weinberg, S. (1977). \emph{The First Three Minutes}. New York: Basic Books. 
  \item Weinberg, S. (2007). Living in the Multiverse. \emph{Universe or Multiverse}, ed.\ B. Carr, pp.\ 29--42.
   Cambridge: Cambridge  University Press. 
\item Weinberg, S. (2008). \emph{Cosmology}. Oxford: Oxford  University Press. 
\item White, R. (2000). Fine-tuning and multiple Universes. \emph{No\^{u}s} 34, 260--276.
\end{trivlist}
\end{small}
\end{document}